\documentclass{jfm}

\usepackage{graphicx}
\usepackage{newtxtext}
\usepackage{newtxmath}
\usepackage{natbib}
\usepackage{hyperref}
\usepackage{bm}
\hypersetup{hidelinks}

\makeatletter
\def\ps@titlepage{%
  \def\@oddhead{}%
  \def\@evenhead{}%
  \def\@oddfoot{\hfil\thepage\hfil}%
  \def\@evenfoot{\hfil\thepage\hfil}%
  \def\sectionmark##1{}%
  \def\subsectionmark##1{}%
}
\def\ps@headings{%
  \def\@oddhead{}%
  \def\@evenhead{}%
  \def\@oddfoot{\hfil\thepage\hfil}%
  \def\@evenfoot{\hfil\thepage\hfil}%
  \def\sectionmark##1{}%
  \def\subsectionmark##1{}%
}
\def\@maketitle#1{%
 \newpage
 \vspace*{10\p@}\addvspace{5pc}%
 {\flushleft
  {\normalfont\LARGE\fontswitch\bfseries
  \put@rapidsHead%
  \@title@alignment%
  \@title \par}%
  \vskip 14\p@ \@plus 2\p@ \@minus 1\p@
  {\normalfont\large\fontswitch\bfseries\baselineskip=12\p@
     \@author@alignment%
     \lowercase{\@author}\par}%
  \vskip 4\p@ \@plus 1\p@
  {\normalfont\small
  \@aff@alignment%
  \@affiliation \par}%
  \par}%
 \vskip 8\p@ \@plus 2\p@ \@minus 1\p@
}
\makeatother

\newcommand{\ii}{{\rm i}}
\newcommand{\e}{{\rm e}}
\newcommand{\cc}{{\rm c.c.}}
\newcommand{\avg}[1]{\left\langle #1 \right\rangle}

\title{A reduced model for surface wave--current interactions without spatial scale separation}

\author{Yohei Onuki\aff{1}, Yasushi Fujiwara\aff{2}}

\affiliation{\aff{1}Research Institute for Applied Mechanics, Kyushu University, Kasuga, Fukuoka 816-8580, Japan
\aff{2}Graduate School of Maritime Sciences, Kobe University, Kobe, Hyogo 658-0022, Japan
}

\begin{document}
\maketitle

\begin{abstract}
We propose a reduced asymptotic model for the mutual interaction between a weakly nonlinear surface gravity wave field and a slowly evolving incompressible current in a homogeneous rotating fluid. The formulation builds on the Craik--Leibovich theory for the wave-averaged momentum equation, but the Stokes drift is not prescribed externally. Instead, it is determined by a companion amplitude equation for a narrow-band wave field concentrated near the wavenumber circle associated with a prescribed carrier frequency. The derivation combines a multiple-time-scale expansion in wave steepness with a phenomenological closure that neglects quartic wave--wave interactions while retaining the third-order Stokes correction. Importantly, no spatial-scale separation is imposed on the wave--current interaction, allowing the wave equation to represent current-induced advection, refraction, and multidirectional scattering. The resulting equations conserve wave action and admit closed energy and momentum budgets for the coupled wave--current system. The model thus provides a tractable bidirectional extension of the classical Craik--Leibovich framework for regimes in which current-induced wave evolution feeds back significantly on the mean flow.
\end{abstract}

\section{Introduction}
Turbulent mixing in the ocean surface boundary layer plays an essential role in the climate system through its influence on sea-surface temperature, air--sea gas exchange, and tracer transport. One of its principal driving mechanisms is Langmuir circulation, arising from the interaction between surface waves and currents \citep{belcher2012global}. The standard theoretical framework for this interaction is the classical Craik--Leibovich (CL) theory \citep{craik1976rational,leibovich1980wave}, in which the wave effect enters the wave-averaged momentum equation through the Stokes drift and the associated vortex force. Because this formulation captures, in a compact way, how surface waves organize the mean-vorticity field into Langmuir circulations, it has provided the foundation for much of the theoretical and numerical literature on Langmuir turbulence and wave-averaged circulation models.

In many applications, the CL equations are closed by prescribing the Stokes drift externally rather than solving for the wave field itself \citep[e.g.,][]{mcwilliams1997langmuir}. This is often an effective approximation when wave evolution can be treated independently of the current dynamics of interest. In that prescribed-wave limit, however, the surface wave field is not an explicit dynamical energy reservoir within the reduced model; instead, it shapes the pathways through which the mean flow is destabilized and rearranged. Actual surface waves, by contrast, are advected, refracted, scattered, and amplitude-modulated by spatially inhomogeneous and time-dependent currents. Such mutual wave--current interactions have recently been reported in direct numerical simulations \citep{fujiwara2020mutual}, ocean current measurements \citep{scully2024vertical}, and linear stability analyses \citep{vergeles2024role,vergeles2026speed}. Surface waves therefore need not be merely catalytic: they may participate actively in energy exchange with Langmuir-type motions.

This viewpoint motivates a wave-resolving closure of the CL framework. Existing two-way coupled wave--current interaction theories provide important precedents. The asymptotic models of \citet{mcwilliams2004asymptotic}, \citet{uchiyama2010wave} and \citet{suzuki2019physical}, together with the wave-action transport model in position--wavevector phase space developed by \citet{vanneste2026consistent}, all go beyond the standard CL closure with prescribed Stokes drift by allowing current-induced refraction and amplitude modulation. Yet these formulations are based either on leading-order single-phase descriptions or on asymptotically scale-separated (i.e., WKB-type) action/ray descriptions. They are therefore not designed to describe, in general form, multidirectional scattering and directional spectral redistribution induced by fully three-dimensional currents such as Langmuir circulations, whose horizontal scales are comparable to the wavelength.

Here we propose a compact model that closes the CL momentum equation by coupling it to an evolution equation for the wave field, without imposing a spatial-scale separation between the waves and the vortical flow. The model is intended for weakly nonlinear, non-breaking surface gravity waves in deep water. Its distinctive spectral assumption is that the wave field is narrowband in frequency around $\omega$, but not localized in propagation direction: the horizontal spectrum is concentrated near the circle $\vert \bm{k}\vert=\kappa$, rather than near a single carrier wavevector. The vortical flow is assumed to be slow, with velocity of the same order as the Stokes drift, and to evolve on the advective time scale associated with that velocity.

Our asymptotic formulation is guided by the reconstituted amplitude equations of \citet{wagner2017asymptotic}, \citet{thomas2017new} and \citet{thomas2018amplitude}, which describe wave propagation through prescribed inhomogeneous environments. The closest analogue to the present bidirectional construction is the reduced model of \citet{xie2015generalised}, in which near-inertial waves are coupled to quasi-geostrophic flows. After deriving asymptotically consistent reduced equation systems, we further introduce phenomenological modifications so that the resulting wave--current system has the desired nonlinear dispersion relation and conservation structure. In particular, the wave equation admits a wave-action invariant, while the coupled system has closed energy and momentum budgets. Together, these properties make the model a closed and energetically consistent extension of the CL framework for wave--current interactions that does not rely on WKB-type scale separation.

The paper is organized as follows. Section~\ref{sec:main_result} states the final reduced equations in dimensional form together with the assumptions under which they are intended to apply. Sections~\ref{sec:equations}--\ref{sec:remodelling} then derive the corresponding nondimensional asymptotic system, from which the equations in Section~\ref{sec:main_result} follow. Specifically, Section~\ref{sec:equations} introduces the dimensional equations and separately scales the wave and current fields, Section~\ref{sec:wave_equation} derives the wave-amplitude equation and the Stokes drift, and Section~\ref{sec:remodelling} modifies the equation system to retain higher-order effects in the wave dispersion relation and improve the energy conservation property. Section~\ref{sec:conservation} establishes the conservation laws for the dimensional reduced system, and Section~\ref{sec:discussion} discusses the limitations of the present closure and suggests possible future work. Appendix~\ref{app:cl_derivation} presents a compact derivation for the CL equation used in the main text.

\section{Reduced model and asymptotic regime} \label{sec:main_result}
We use Cartesian coordinates $(x_1,x_2,z)$, with $z$ directed upward, and consider an incompressible fluid in $-h \le z \le 0$ with a flat bottom and a free surface. Note that horizontal indices $i,j$ run over $1,2$ and are summed when repeated. The averaged flow dynamics involves two kinds of velocities: the Lagrangian-mean velocity $\bm{U}^L=(U^L_1, U^L_2, W^L)$ and the horizontal Stokes drift $\bm{U}^s = (U^s_1, U^s_2, 0)$. We impose the incompressible constraint and the impermeable boundary conditions for the Lagrangian-mean velocity,
\begin{align}
\nabla \cdot \bm{U}^L =0, \qquad W^L=0 \quad \text{at} \quad z=-h,0 .
\end{align}
We adopt the traditional approximation for rotation: the Coriolis parameter $f$ is taken constant and only the vertical component of the planetary rotation enters the equations. Here and below, $\bm{z}$ denotes the upward unit vector, and $t$ denotes the time variable. The wave-averaged momentum equation is given by
\begin{align} \label{eq:summary_mean_momentum}
\bm{U}^L_t + \bm{U}^L \cdot \nabla \bm{U}^L + (f \bm{z} - \nabla \times \bm{U}^s) \times \bm{U}^L
= - \nabla \Pi + \bm{U}^s_t ,
\end{align}
where $\Pi$ is a modified pressure enforcing the incompressible constraint. This is a variant of the familiar CL equation \citep{suzuki2016understanding}.

The wave field is assumed to be narrow-band in intrinsic frequency and concentrated near the circle $\vert \bm{k}\vert=\kappa$ in horizontal wavenumber space, without imposing a preferred propagation direction. We propose an evolution equation for the wave field specified by a complex amplitude $A(x_1,x_2,t)$ that obeys
\begin{align} \label{eq:summary_amplitude}
\left[1+\alpha\left(\partial_i^2+\kappa^2\right)\right] A_t
- \frac{\ii \omega_\kappa}{2 \kappa}\left(\partial_i^2+\kappa^2\right)A
= \frac{\omega_\kappa}{2\kappa \omega} \mathcal{L}(\bm{U}^L, A) ,
\end{align}
where
\begin{subequations} \label{eq:summary_operator}
\begin{align}
\mathcal{L} (\bm{U}^L, A) & \equiv
\left(\overline{U}^L_i A_{,ij} \right)_{,j}
+ \left(\overline{U}^L_i A_{,j} \right)_{,ij}
- \kappa^2 \tilde{U}^L_i A_{,i}
- \kappa^2 \left(\tilde{U}^L_i A\right)_{,i}, \\
\overline{U}^L_i &\equiv \int_{-h}^0 \Phi^2 U^L_i \, dz, \qquad
\tilde{U}^L_i \equiv \frac{1}{\kappa^2}\int_{-h}^0 \Phi_z^2 U^L_i \, dz .
\end{align}
\end{subequations}
The coefficient
\begin{align}
\alpha = \frac{\kappa \omega_{\kappa\kappa}-\omega_\kappa}{4\kappa^2\omega_\kappa}
\end{align}
is the standard reconstitution parameter used to improve the linear dispersion accuracy \citep{thomas2018amplitude}. The intrinsic frequency satisfies the linear finite-depth dispersion relation
\begin{align}
\omega^2 = g \kappa \tanh (\kappa h) ,
\end{align}
and $\Phi(z)$ is the normalized vertical structure function
\begin{align} \label{eq:vertical_structure_function}
\Phi(z) = \frac{C \cosh\{\kappa(z+h)\}}{\cosh(\kappa h)}, \qquad
\int_{-h}^0 \Phi^2\,dz = 1,
\end{align}
with $C^2 = g \kappa / (\omega \omega_\kappa)$.

The Stokes drift appearing in \eqref{eq:summary_mean_momentum} is computed from $A$ as
\begin{align} \label{eq:summary_stokes}
U^s_i &= \frac{-\ii}{\omega}
\left(
\Phi^2 A^\dagger_{,j} A_{,ij}
+ \Phi_z^2 A^\dagger A_{,i}
\right)+\cc ,
\end{align}
where ${}^\dagger$ denotes complex conjugation.

The model is intended for weakly nonlinear, non-breaking waves and slowly evolving vortical currents in the following asymptotic regime.
\begin{enumerate}
\item The wave steepness is small, measured by $\epsilon \ll 1$.
\item The mean-current velocity is $O(\epsilon)$ compared with the wave orbital velocity, so the wave and mean-flow time scales differ by a factor $\epsilon^{-2}$.
\item The wave field is narrow-band in intrinsic frequency around $\omega$, but not in propagation direction: its horizontal wavenumber support is concentrated near $\vert \bm{k}\vert=\kappa$.
\item Quartic wave--wave interactions are neglected in a quasi-linear closure, while current-induced wave evolution and an approximate Stokes correction for deep-water waves are retained.
\end{enumerate}
The following sections present how these assumptions lead to the aforementioned reduced equation system from the unaveraged momentum equation.

\section{Governing equations and scaling} \label{sec:equations}
We begin from the dimensional incompressible Euler equations in a rotating frame. In this section, dimensional variables are marked by an asterisk; after nondimensionalization, the asterisk is dropped. The governing equations are
\begin{align}
\bm{u}^\ast_{t^\ast} + \bm{u}^\ast \cdot \nabla^\ast \bm{u}^\ast + f^\ast \bm{z} \times \bm{u}^\ast
= - \nabla^\ast p^\ast - g^\ast \bm{z},
\end{align}
where $\bm{u}^\ast=(u_1^\ast,u_2^\ast,w^\ast)$ and $\nabla^\ast \cdot \bm{u}^\ast=0$. The free-surface and bottom boundary conditions are
\begin{subequations}
\begin{align}
\zeta^\ast_{t^\ast} + u_i^\ast \zeta^\ast_{,i} &= w^\ast,
\quad z^\ast=\zeta^\ast(x_1^\ast,x_2^\ast,t^\ast), \\
p^\ast &= 0,
\quad z^\ast=\zeta^\ast(x_1^\ast,x_2^\ast,t^\ast), \\
w^\ast &= 0,
\quad z^\ast=-h^\ast .
\end{align}
\end{subequations}

\subsection{Separation into wave and vortical components}
We separate the flow into a fast wave part and a slow vortical part,
\begin{align}
\bm{u}^\ast = \bm{u}^{\prime \ast} + \bm{U}^\ast, \qquad
p^\ast = -g^\ast z^\ast + p^{\prime \ast} + P^\ast, \qquad
\zeta^\ast = \zeta^{\prime \ast} + Z^\ast .
\end{align}
Primes denote oscillatory wave quantities, while uppercase symbols denote the slowly evolving mean flow.

The water depth $h^\ast$ is used to nondimensionalize space, so that $\bm{x}^\ast=h^\ast \bm{x}$. Let $\epsilon \ll 1$ denote wave steepness. We scale the wave variables as
\begin{align}
\zeta^{\prime \ast} = \epsilon h^\ast \zeta', \qquad
\bm{u}^{\prime \ast} = \epsilon \sqrt{g^\ast h^\ast}\,\bm{u}', \qquad
p^{\prime \ast} = \epsilon g^\ast h^\ast p' .
\end{align}
The mean flow is assumed comparable to the Stokes drift, hence
\begin{align}
\bm{U}^\ast = \epsilon^2 \sqrt{g^\ast h^\ast}\,\bm{U}, \qquad
P^\ast = \epsilon^4 g^\ast h^\ast P, \qquad
Z^\ast = \epsilon^4 h^\ast Z .
\end{align}
The fast wave time scale is $(h^\ast/g^\ast)^{1/2}$ and the slow mean-flow time scale is $\epsilon^{-2}(h^\ast/g^\ast)^{1/2}$. We therefore introduce the slow time
\begin{align}
T = \epsilon^2 t,
\end{align}
and scale the Coriolis parameter so that rotational effects enter the mean dynamics at leading order,
\begin{align}
f^\ast = \epsilon^2 \sqrt{g^\ast/h^\ast}\,f .
\end{align}
With these scalings, the governing equations become
\begin{subequations} \label{eq:scaled_governing_equations}
\begin{align}
\bm{u}_t + \epsilon \bm{u}\cdot\nabla \bm{u} + \epsilon^2 f \bm{z}\times \bm{u}
&= - \nabla p,
\quad -1 \le z \le \epsilon \zeta, \label{eq:scaled_momentum_equation} \\
\zeta_t + \epsilon u_i \zeta_{,i} &= w,
\quad z=\epsilon \zeta, \\
p &= \zeta,
\quad z=\epsilon \zeta, \\
w &= 0,
\quad z=-1,
\end{align}
\end{subequations}
with
\begin{align}
\bm{u} = \bm{u}' + \epsilon \bm{U}, \qquad
p = p' + \epsilon^3 P, \qquad
\zeta = \zeta' + \epsilon^3 Z 
\end{align} 
The oscillatory variables depend on $(\bm{x},t,T)$, with $t$ and $T$ treated as independent multiple-scale times, whereas $(\bm{U},P,Z)$ depend on $(\bm{x},T)$.

\subsection{Temporal averaging and the mean-momentum equation}
We denote by $\avg{\cdot}$ an average over times long compared with the wave period but short compared with the slow mean-flow evolution. The wave variables have zero average, so that $\avg{\bm{u}}=\epsilon \bm{U}$, but quadratic wave correlations generally do not vanish. A direct average of \eqref{eq:scaled_governing_equations} leaves the wave contribution hidden inside the quadratic term $\epsilon \avg{\bm{u} \cdot \nabla \bm{u}}$. Following the seminal work of \citet{craik1976rational}, one rewrites the nonlinearity in vector-invariant form and uses the $O(\epsilon^2)$ wave-vorticity equation to reduce that average to a Stokes-drift/mean-vorticity coupling. The details are given in Appendix~\ref{app:cl_derivation}. The resulting wave-averaged momentum equation is
\begin{align} \label{eq:averaged_momentum_with_error}
\bm{U}^L_T + \bm{U}^L \cdot \nabla \bm{U}^L + (f \bm{z} - \nabla \times \bm{U}^s) \times \bm{U}^L
= - \nabla \Pi + \bm{U}^s_T
+ O(\epsilon),
\end{align}
where $\bm{U}^L = \bm{U} + \bm{U}^s$ with $\bm{U}^s$ the Stokes drift, and $\Pi$ collects the mean pressure and several wave-induced gradient terms generated in the Appendix. A central task in the remaining steps is to obtain a consistent expression for $\bm{U}^s$ from the wave dynamics.

\section{Wave equation} \label{sec:wave_equation}
To derive the wave evolution, we expand the wave part in powers of $\epsilon$,
\begin{align}
(\bm{u}',p') = (\bm{u}_0,p_0) + \epsilon (\bm{u}_1,p_1) + \epsilon^2 (\bm{u}_2,p_2) + \cdots .
\end{align}
Within the quasi-linear approximation adopted here, quadratic products of wave variables are neglected in the wave equation, which removes the first-order terms from the expansion. This closure isolates the current-induced modulation of the narrow-band wave field and leaves quartic resonant wave--wave interactions outside the scope of the present consideration.

\subsection{Leading-order wave field}
At $O(1)$, the equations reduce to
\begin{subequations}
\begin{align}
\partial_t \bm{u}_0 &= - \nabla p_0,
\quad -1 \le z \le 0, \\
\partial_t p_0 &= w_0,
\quad z=0, \\
w_0 &= 0,
\quad z=-1 
\end{align}
\end{subequations}
with $\nabla \cdot \bm{u}_0 = 0$. The leading-order wave motion is irrotational, so we may introduce a flow potential to write $\bm{u}_0=\nabla \phi_0$ with $\phi_{0t}=-p_0$. We seek a time-harmonic solution
\begin{align}
\phi_0 = \e^{-\ii \omega t}\hat{\phi}_0 + \cc .
\end{align}
Then $\hat{\phi}_0$ satisfies
\begin{subequations}
\begin{align}
\nabla^2 \hat{\phi}_0 &= 0,
\quad -1 \le z \le 0, \\
\hat{\phi}_{0z} - \omega^2 \hat{\phi}_0 &= 0,
\quad z=0, \\
\hat{\phi}_{0z} &= 0,
\quad z=-1 .
\end{align}
\end{subequations}
Defining $\kappa$ through $\omega^2=\kappa\tanh\kappa$, we may write a solution as
\begin{align} \label{eq:phi_separate_form}
\hat{\phi}_0 = \Phi(z) A_0(x_1,x_2,T),
\end{align}
where $\Phi$ is the vertical structure function defined by \eqref{eq:vertical_structure_function} whose dimensionless form is $\Phi = C \cosh \{ \kappa (z + 1) \} / \cosh \kappa$. The horizontal structure function $A_0$ obeys the Helmholtz equation
\begin{align} \label{eq:helmholtz_A0}
(\partial_i^2+\kappa^2)A_0 = 0 .
\end{align}
The normalization condition of $\Phi$ enables \eqref{eq:phi_separate_form} to be rewritten as
\begin{align} \label{eq:A0_p0_relation}
A_0 = \int_{-1}^0 \Phi \hat{\phi}_0\,dz
= \left< \int_{-1}^0 \frac{\e^{\ii \omega t}\Phi p_0}{\ii \omega}\,dz \right> ,
\end{align}
where we have used $\left< \e^{2 \ii \omega t} \right> = 0$. The expressions of $p_0$ and $w_0$ at $z=0$ in terms of $A_0$ are also worth noting:
\begin{align}
p_0 = \ii \omega C A_0 \e^{-\ii \omega t} + \cc,
\qquad
w_0 = C \omega^2 A_0 \e^{-\ii \omega t} + \cc ,
\end{align}
which will be used in the next step.

\subsection{Second-order solvability condition}
At $O(\epsilon^2)$, the wave equations become
\begin{subequations} \label{eq:order2_wave_momentum}
\begin{align}
\bm{u}_{2t} + \bm{u}_{0T} + \bm{U}\cdot\nabla \bm{u}_0 + \bm{u}_0\cdot\nabla \bm{U}
+ f \bm{z}\times \bm{u}_0
&= - \nabla p_2,
\quad -1 \le z \le 0, \label{eq:order2_wave_momentum_a}\\
p_{2t} + p_{0T} + U_i p_{0,i}
&= w_2 + W_z p_0,
\quad z=0, \label{eq:order2_wave_momentum_b}\\
p_{2z} &= 0,
\quad z=-1 \label{eq:order2_wave_momentum_c}
\end{align}
\end{subequations}
with $\nabla \cdot \bm{u}_2 = 0$. Taking the divergence of \eqref{eq:order2_wave_momentum_a} and using $\nabla\cdot\bm{U}=0$ and $\nabla\times\bm{u}_0=0$ gives
\begin{align} \label{eq:p2_poisson}
\nabla^2 p_2 = - 2 \nabla \cdot (\bm{U}\cdot\nabla \bm{u}_0) .
\end{align}
To isolate the resonant part of the $O(\epsilon^2)$ problem, we write
\begin{align}
(\bm{u}_2,p_2) = (\hat{\bm{u}}_2,\hat{p}_2)\e^{-\ii \omega t} + \cc + \mathcal{R},
\end{align}
where $\mathcal{R}$ contains non-resonant terms.

Now, we use three expressions derived so far: the vertical component of \eqref{eq:order2_wave_momentum_a} evaluated at $z=0$, the free-surface condition \eqref{eq:order2_wave_momentum_b}, and \eqref{eq:p2_poisson} multiplied by $\Phi$ and integrated over $-1 \le z \le 0$. Multiplying these by $\e^{\ii \omega t}$ and applying the temporal average removes non-resonant terms. The resulting equations are
\begin{subequations} \label{eq:resonant_system}
\begin{align}
-\ii \omega \hat{w}_2 + \hat{p}_{2z}
&= - C \omega^2 A_{0T}
- C \omega^2 \left(U_i A_{0,i} - U_{i,i}A_0\right) , \label{eq:resonant_system_a}\\
-\ii \omega \hat{p}_2 - \hat{w}_2
&= - \ii \omega C A_{0T} - \ii \omega C (U_i A_0)_{,i} , \label{eq:resonant_system_b}\\
C(\hat{p}_{2z}-\omega^2 \hat{p}_2)
&= - \ii \omega (\partial_i^2+\kappa^2)A_2
- \mathcal{L} (\bm{U}, A_0)
- 2 C^2 \omega^2 U_i A_{0,i}, \label{eq:resonant_system_c}
\end{align}
\end{subequations}
where $\hat{p}_2$, $\hat{w}_2$, and $U_i$ are evaluated at $z=0$, the second-order wave amplitude is defined following \eqref{eq:A0_p0_relation} as
\begin{align}
A_2 \equiv \avg{\int_{-1}^0 \frac{\e^{\ii \omega t}\Phi p_2}{\ii \omega}\,dz},
\end{align}
and $\mathcal{L}$ is a bilinear differential operator whose explicit form is given in \eqref{eq:summary_operator}. Eliminating $\hat{w}_2$ and $\hat{p}_2$ from \eqref{eq:resonant_system} then yields the solvability condition
\begin{align} \label{eq:A0T}
A_{0T} - \frac{\ii \omega_\kappa}{2\kappa}(\partial_i^2+\kappa^2)A_2
= \frac{\omega_\kappa}{2\kappa \omega} \mathcal{L}(\bm{U}, A_0) .
\end{align}
Reconstituting the wave amplitude to second order,
\begin{align}
A \equiv \avg{\int_{-1}^0 \frac{\e^{\ii \omega t}\Phi p'}{\ii \omega}\,dz}
= A_0 + \epsilon^2 A_2 + O(\epsilon^3),
\end{align}
and combining \eqref{eq:A0T} with \eqref{eq:helmholtz_A0}, we obtain an amplitude equation on the slow time scale,
\begin{align} \label{eq:amplitude_basic}
A_T - \frac{\ii \omega_\kappa}{2\epsilon^2 \kappa}(\partial_i^2+\kappa^2)A
= \frac{\omega_\kappa}{2\kappa \omega} \mathcal{L}(\bm{U}, A) + O(\epsilon).
\end{align}

\subsection{Stokes drift} \label{sec:stokes_drift}
The leading-order expressions for the velocity and particle displacement associated with the wave field are given as
\begin{subequations}
\begin{align}
\bm{u}' &= \nabla (A\Phi)\e^{-\ii \omega t} + \cc + O(\epsilon), \\
\bm{\xi}' &= \frac{\ii \nabla (A\Phi)\e^{-\ii \omega t}}{\omega} + \cc + O(\epsilon) ,
\end{align}
\end{subequations}
where we have used $\bm{u}' = \partial_t \bm{\xi}' + \mathcal{O}(\epsilon)$. The Stokes drift is consequently given by
\begin{align} \label{eq:Stokes_with_error}
\bm{U}^s = \avg{\bm{\xi}'\cdot\nabla \bm{u}'}
= \left( U^s_1, U^s_2, 0 \right) + O(\epsilon),
\end{align}
with the components $(U^s_1, U^s_2)$ defined as \eqref{eq:summary_stokes}. In this form, at the retained order, the Stokes drift is obtained explicitly as a quadratic functional of the wave amplitude.

\section{Remodelling the equations} \label{sec:remodelling}

We have developed a closed equation system consisting of the mean momentum equation \eqref{eq:averaged_momentum_with_error}, wave amplitude equation \eqref{eq:amplitude_basic}, and the explicit expression of the Stokes drift \eqref{eq:Stokes_with_error} with \eqref{eq:summary_stokes}. The purpose of this section is to \textit{remodel} these equations by recovering several higher-order effects that were neglected so far. The resulting modified equation system has an improved wave dispersion relation and clearer conservation properties.

\subsection{Linear wave dispersion relation}
The wave equation \eqref{eq:amplitude_basic} has the correct retained-order coupling structure, but its linear dispersion relation is only the low-order asymptotic approximation inherited from the multiple-scale expansion. A standard improvement is therefore to add a formally higher-order time-derivative term that leaves the asymptotic accuracy of the nonlinear dynamics unchanged while yielding a more accurate linear dispersion relation and a better-behaved reduced model when the wave field develops shorter horizontal scales. Using $(\partial_i^2+\kappa^2)A_T = O(\epsilon^2)$ from \eqref{eq:helmholtz_A0}, one may add $\alpha(\partial_i^2+\kappa^2)A_T$ without changing the retained order. Choosing
\begin{align}
\alpha = \frac{\kappa \omega_{\kappa\kappa}-\omega_\kappa}{4\kappa^2\omega_\kappa}
\end{align}
improves the linear dispersion relation to higher order \citep{thomas2018amplitude}.

\subsection{Incompressibility condition}

In their original definitions, the Eulerian mean velocity $\bm{U}$ is divergence-free. As for the Stokes drift $\bm{U}^s$, if the leading Helmholtz constraint $(\partial_i^2 + \kappa^2) A = 0$ were exactly satisfied, then $\nabla \cdot \bm{U}^{s}=0$ would follow. However, in the reconstituted amplitude equation, this holds only up to $O(\epsilon^2)$. As a result, a Lagrangian mean velocity defined by $\bm{U}^L = \bm{U} + \bm{U}^s$ would not be divergence-free.

Here, we rectify this situation by replacing the incompressibility condition on $\bm{U}$ with the incompressibility of the Lagrangian mean velocity $\bm{U}^L$. This modification does not influence the asymptotic accuracy because the modified divergent part is $\mathcal{O}(\epsilon^2)$. The motivation for introducing the divergence-free correction for $\bm{U}^L$ is to make the coupled wave--current system satisfy the energy conservation law established in Section~\ref{sec:conservation}. Without this modification, a flow divergence would introduce spurious source terms into the energy budget.

\subsection{Stokes correction} \label{sec:stokes_correction}

In the framework of the quasi-linear closure, because wave self-interactions are ignored, the Stokes correction to the dispersion relation has been excluded. Here, we partly remedy this issue by replacing $\mathcal{L}(\bm{U}, A)$ on the right-hand side of the amplitude equation \eqref{eq:amplitude_basic} with $\mathcal{L}(\bm{U}, A) + \mathcal{L}(\bm{U}^s, A) = \mathcal{L}(\bm{U}^L, A)$. As a result, the modified wave--current interaction term, involving the Stokes drift, provides a good approximation to the third-order Stokes correction to the wave frequency in deep water. Note that this modification is not derived from the present asymptotics but was introduced phenomenologically following the recent work of \citet{vanneste2026consistent}. For shallow-water conditions, the present model may not appropriately represent the Stokes correction, and therefore we should instead directly account for wave--wave interaction asymptotics.

In the end, after applying these remodelling procedures, restoring dimensions and suppressing the bookkeeping parameter $\epsilon$, we obtain the reduced dimensional system stated in Section~\ref{sec:main_result}.

\section{Conservation properties} \label{sec:conservation}
The proposed reduced system possesses several conservation laws. The wave equation conserves wave action, while the coupled system conserves an energy invariant representing wave--current energy exchange. In the absence of rotation, the system also conserves the sum of the Eulerian-mean momentum and the wave pseudomomentum. This section presents the derivation of these invariants. For simplicity, we assume that the horizontal domain is periodic or unbounded so that, after spatial integration, the horizontal flux terms always vanish.

\subsection{Wave action}
Multiplying \eqref{eq:summary_amplitude} by $A^\dagger$, adding the complex conjugate and integrating over the horizontal plane gives
\begin{align}
\frac{d}{dt}\iint
\left[
\vert A\vert^2
- \alpha\left(\vert A_{,i}\vert^2-\kappa^2\vert A\vert^2\right)
\right]\,dx\,dy
= 0 .
\end{align}
The conserved functional thus derived,
\begin{align} \label{eq:action_invariant}
\mathcal{A} \equiv
\iint
\left[
\vert A\vert^2
- \alpha\left(\vert A_{,i}\vert^2-\kappa^2\vert A\vert^2\right)
\right] dx dy ,
\end{align}
coincides with the first invariant identified by \citet{thomas2018amplitude}, specifically Eq.~(24a) in their paper. We regard this $\mathcal{A}$ as the wave action in the present model since its conservation follows from a separation of time scales between waves and currents---a fundamental requirement for the existence of an adiabatic invariant.

\subsection{Energy}
Next, multiplying \eqref{eq:summary_amplitude} by $A_t^\dagger$, subtracting the complex conjugate and integrating horizontally yields
\begin{align} \label{eq:wave_energy_balance}
\frac{d}{dt}
\iint
\left(
\vert A_{,i}\vert^2-\kappa^2\vert A\vert^2
\right)\,dx\,dy
= - \iiint \bm{U}^L \cdot \bm{U}^s_t dx dy dz .
\end{align}
Dotting \eqref{eq:summary_mean_momentum} with $\bm{U}^L$, integrating over the full domain, and using $\nabla\cdot\bm{U}^L=0$ together with $W^L=0$ at the boundaries gives
\begin{align} \label{eq:mean_energy_balance}
\frac{d}{dt}\iiint \frac{\left\vert \bm{U}^L \right\vert^2}{2} dx dy dz
= \iiint \bm{U}^L \cdot \bm{U}^s_t dx dy dz .
\end{align}
Combining \eqref{eq:wave_energy_balance} and \eqref{eq:mean_energy_balance}, we find an energy invariant in the coupled wave--current system
\begin{align} \label{eq:energy}
\mathcal{E} \equiv
\iint
\left(
\vert A_{,i}\vert^2-\kappa^2\vert A\vert^2
\right) dx dy
+ \iiint \frac{\left\vert \bm{U}^L \right\vert^2}{2} dx dy dz .
\end{align}
In the original inviscid Craik--Leibovich equation, the second term in \eqref{eq:energy} is conserved for a prescribed time-independent Stokes drift $\bm{U}^s$. The present model extends this classical result by incorporating energetic feedback to the wave field in response to the growth or decay of the Lagrangian-mean current.

It is important to note that \eqref{eq:energy} is not the total energy of the full system. In the present model, most of the physical energy resides in the wave field and is proportional to the wave action $\mathcal{A}$. Only an $\mathcal{O}(\epsilon^2)$ fraction of the energy, represented by $\mathcal{E}$, is exchanged between waves and currents. Accordingly, a linear combination of $\mathcal{A}$ and $\mathcal{E}$ reflects total energy conservation. This property resembles that of the coupled model of near-inertial waves and quasi-geostrophic flows proposed by \citet{xie2015generalised}.

\subsection{Momentum}

We finally show that, when the Coriolis parameter is set to zero, the coupled system also conserves horizontal momentum---the sum of the wave pseudomomentum and the mean-flow momentum. To derive this conservation law, one should not identify the pseudomomentum with the vertical integral of the Stokes drift. There is an $\mathcal{O}(\epsilon^2)$ mismatch owing to the reconstitution procedure.

We first derive the wave contribution. Multiplying \eqref{eq:summary_amplitude} by $A^\dagger_{,i}$, subtracting the complex conjugate equation multiplied by $A_{,i}$, and integrating over the horizontal plane gives, after integration by parts,
\begin{align} \label{eq:wave_pseudomomentum_balance}
\frac{d \mathcal{P}^{\rm w}_i}{dt}
= - \iiint U^s_j U^L_{j,i}\,dx\,dy\,dz ,
\end{align}
with the pseudomomentum defined as
\begin{align}
\mathcal{P}^{\rm w}_i
\equiv \frac{\kappa}{\ii\omega_\kappa} \iint
\left[ (1+\alpha\kappa^2) \left( A^\dagger A_{,i} - A A^\dagger_{,i} \right)
- \alpha \left( A^\dagger_{,j}A_{,ij} - A_{,j}A^\dagger_{,ij} \right) \right] dx dy .
\label{eq:wave_pseudomomentum}
\end{align}
Thus the pseudomomentum changes through the refraction-like exchange term involving the Stokes drift and the horizontal gradient of the Lagrangian-mean velocity.

The corresponding mean-flow balance is obtained by rewriting the mean momentum equation in a conservative form. The procedure is essentially the same as that of \citet{vanneste2026consistent}. For $f=0$, the horizontal components of \eqref{eq:summary_mean_momentum} may be rewritten as
\begin{align}
\frac{\partial U_i}{\partial t} + \nabla \cdot \left( U_i \bm{U}^L \right)
- U^s_j U^L_{j,i} + \left( \Pi + \bm{U}^L \cdot \bm{U}^s \right)_{,i} = 0 .
\label{eq:eulerian_momentum_local_balance}
\end{align}
(Note that the Eulerian-mean velocity $\bm{U} = \bm{U}^L - \bm{U}^s$ is not divergence-free under the present definition.) Integrating \eqref{eq:eulerian_momentum_local_balance} over the full fluid domain gives
\begin{align}
\frac{d}{dt} \iiint U_i \,dx\,dy\,dz = \iiint U^s_j U^L_{j,i}\,dx\,dy\,dz .
\label{eq:eulerian_momentum_balance}
\end{align}
Combining \eqref{eq:wave_pseudomomentum_balance} and \eqref{eq:eulerian_momentum_balance} yields a momentum invariant
\begin{align}
\mathcal{P}_i \equiv \mathcal{P}^{\rm w}_i + \iiint U_i \,dx\,dy\,dz .
\end{align}
Thus, in the non-rotating case, the reduced model conserves the sum of the Eulerian-mean momentum and the wave pseudomomentum.

If the wave field strictly satisfies the Helmholtz equation $(\partial_i^2 + \kappa^2) A = 0$, the pseudomomentum \eqref{eq:wave_pseudomomentum} reduces to the spatial integral of the Stokes drift. In that special case, the total momentum $\mathcal{P}_i$ coincides with $\iiint U^L_i \,dx\,dy\,dz$, thus agreeing with earlier results of \citet{vanneste2026consistent}.

\section{Summary and future work} \label{sec:discussion}

We have developed a reduced model for interactions between surface gravity waves and a mean current. The model closes the Craik--Leibovich mean-momentum equation by coupling it to a physical-space amplitude equation, so that the Stokes drift is determined dynamically rather than prescribed externally. The formulation assumes weak nonlinearity, constant depth and a narrow frequency band, but does not impose a spatial-scale separation between the waves and the current. It can therefore represent current-induced advection, refraction and multidirectional scattering of the wave field in physical space. The resulting coupled system conserves wave action and admits consistent energy and momentum budgets.

While we have considered an inviscid homogeneous fluid, the present model can be directly extended to a viscous Boussinesq flow model. Assuming that the viscosity and buoyancy terms in the original Navier--Stokes system are small, so that the leading-order wave dynamics are unchanged, we may readily incorporate these effects into the averaged CL equation \eqref{eq:summary_mean_momentum}. To retain asymptotic consistency, we may also need to add linear correction terms to the wave amplitude equation \eqref{eq:summary_amplitude}.

A complementary next step is to implement the reduced equations in a numerical model and use them to examine the energy exchange between surface waves and Langmuir circulations. Such calculations would provide a direct test of the main physical mechanism isolated in this paper: the Stokes drift is not an externally prescribed forcing, but evolves with the wave amplitude and can therefore act as part of a dynamical energy reservoir. The conservative structure derived in Section~\ref{sec:conservation} provides a natural diagnostic for separating reversible wave--current energy transfer from numerical dissipation or from additional physical parameterisations introduced in more realistic models. Comparing such results with otherwise identical CL simulations using a prescribed Stokes drift would clarify when the bidirectional wave evolution retained here materially changes the growth, saturation and energetics of Langmuir circulations.

Other extensions are more substantial. The most immediate is to restore wave--wave interactions and derive the corresponding cubic nonlinear terms in \eqref{eq:summary_amplitude} instead of the ad~hoc Stokes-drift corrections introduced in Section~\ref{sec:stokes_correction}. This would be done via standard multiple-time-scale asymptotics, but it remains unsettled whether action and energy conservation can be retained in such an extended model. Further challenges include bathymetric variations and mean sea-surface elevations, wave breaking, and coupling to atmospheric forcing. Each of these would move the model closer to upper-ocean applications while preserving the central idea of solving for the wave field and the mean flow simultaneously.

\acknowledgments
This study was supported by KAKENHI Grant Number JP20K14556, JP26K00783 and by the Collaborative Research Program of Research Institute for Applied Mechanics, Kyushu University. The authors used GPT-5.5 (OpenAI) to assist with improving the clarity and readability of the manuscript. All scientific content remains the responsibility of the authors. 

\appendix

\section{Derivation of the Craik--Leibovich mean-momentum equation} \label{app:cl_derivation}
This Appendix gives a compact derivation of the CL mean-momentum equation \eqref{eq:averaged_momentum_with_error}. Although this result is well established \citep[e.g.,][]{leibovich1977convective,leibovich1980wave}, we present it here for completeness and to clarify the assumptions and approximations involved.

It is convenient to write the interior momentum equation \eqref{eq:scaled_momentum_equation} as
\begin{align} \label{app:vector_invariant_exact}
\bm{u}_t
- \epsilon \bm{u}\times\left(\nabla\times\bm{u}+\epsilon f \bm{z}\right)
+ \nabla \pi
= \bm{0},
\qquad
\pi \equiv p + \frac{\epsilon}{2}\vert \bm{u}\vert^2 .
\end{align}
Since the leading-order wave motion is irrotational, the absolute vorticity can be represented as
\begin{align}
\nabla\times\bm{u} + \epsilon f \bm{z} &= \epsilon \bm{\Omega}^a + \epsilon^2 \bm{\omega}_2 + O(\epsilon^3),
\end{align}
with
\begin{align}
\bm{\Omega}^a \equiv \nabla \times \bm{U} + f \bm{z}, \qquad \bm{\omega}_2 \equiv \nabla \times \bm{u}_2 .
\end{align}
Note that the leading-order vorticity $\bm{\Omega}^a$ is a function of the slow time variable $T = \epsilon^2 t$ while the next-order vorticity $\bm{\omega}_2$ depends on $t$. Averaging \eqref{app:vector_invariant_exact} over the fast wave time gives
\begin{align} \label{app:mean_vortex_balance}
\bm{U}_T
- \bm{U} \times \bm{\Omega}^a
- \avg{\bm{u}_0\times\bm{\omega}_2}
+ \nabla \Pi_0
= O(\epsilon),
\end{align}
for a suitably rescaled mean pressure $\Pi_0$.

The remaining task is to evaluate the correlation term $\left< \bm{u}_0 \times \bm{\omega}_2 \right>$. Taking the curl of \eqref{app:vector_invariant_exact} gives
\begin{align} \label{app:omega2_transport}
\bm{\omega}_{2t}
= \nabla \times \left( \bm{u}_0 \times \bm{\Omega}^a \right) .
\end{align}
Let $\bm{\xi}_0$ be the leading particle displacement, defined by $\partial_t \bm{\xi}_0 = \bm{u}_0$. Integration of \eqref{app:omega2_transport} in time yields
\begin{align} \label{app:omega2_from_xi}
\bm{\omega}_2
= \nabla\times\left(\bm{\xi}_0\times\bm{\Omega}^a\right) .
\end{align}
This is the Craik--Leibovich wave-vorticity estimate: the wave's leading orbital motion tilts and stretches the slowly varying mean absolute vorticity.

Using an identity for solenoidal vector fields, we derive
\begin{align}
& \bm{u}_0 \times \bm{\omega}_2 = \bm{u}_0 \times \left[ \nabla\times\left(\bm{\xi}_0\times\bm{\Omega}^a\right) \right] \nonumber \\
= & \nabla\left[\bm{u}_0\cdot(\bm{\xi}_0\times\bm{\Omega}^a)\right]
+ \left[(\bm{\xi}_0\cdot\nabla)\bm{u}_0 - (\bm{u}_0\cdot\nabla)\bm{\xi}_0\right]\times\bm{\Omega}^a 
+ \bm{\xi}_0 \times \left[ \nabla \times \left( \bm{u}_0 \times \bm{\Omega}^a \right) \right] .
\end{align}
Averaging over the fast time gives
\begin{align} \label{app:u0_w2}
\avg{\bm{u}_0\times\bm{\omega}_2}
&=
\nabla \chi
+ \left(
\avg{(\bm{\xi}_0\cdot\nabla)\bm{u}_0}
-
\avg{(\bm{u}_0\cdot\nabla)\bm{\xi}_0}
\right)\times\bm{\Omega}^a
+ \avg{\bm{\xi}_0\times\bm{\omega}_{2t}},
\end{align}
with $\chi \equiv \avg{\bm{u}_0\cdot(\bm{\xi}_0\times\bm{\Omega}^a)}$. Now, using the identity $\left< \partial_t a \right> = 0$ for any variable $a$, we derive the formulas
\begin{align}
\avg{(\bm{u}_0\cdot\nabla)\bm{\xi}_0}
= - \avg{(\bm{\xi}_0\cdot\nabla)\bm{u}_0} = - \bm{U}^s + O(\epsilon),
\qquad
\avg{\bm{\xi}_0\times\bm{\omega}_{2t}}
= - \avg{\bm{u}_0\times\bm{\omega}_2} 
\end{align}
and hence rewrite \eqref{app:u0_w2} as
\begin{align} \label{app:wave_vortex_force}
\avg{\bm{u}_0\times\bm{\omega}_2}
=
\bm{U}^s \times\bm{\Omega}^a
+ \nabla\left(\frac{\chi}{2}\right) + O (\epsilon).
\end{align}
Substituting \eqref{app:wave_vortex_force} into \eqref{app:mean_vortex_balance} and performing some manipulations recovers the CL momentum equation \eqref{eq:averaged_momentum_with_error}.

\bibliographystyle{jfm}
\bibliography{jfm}

\end{document}